\documentclass[10pt, conference, letterpaper]{IEEEtran}
\IEEEoverridecommandlockouts
\usepackage{cite}
\usepackage{amsmath,amssymb,amsfonts}
\usepackage{graphicx}
\usepackage[caption=false,font=footnotesize]{subfig}
\usepackage{textcomp}
\usepackage{xcolor}

\usepackage{tabularx}
\usepackage{multirow}
\usepackage{diagbox}
\usepackage[binary-units]{siunitx} 
\DeclareSIUnit{\bps}{bps}
\usepackage{amsthm}

\begin{document}

\title{Unified Modeling and Performance Comparison for Cellular and Cell-Free Massive MIMO}

\author{\IEEEauthorblockN{Wei Jiang\IEEEauthorrefmark{1} and Hans D. Schotten\IEEEauthorrefmark{2}}
\IEEEauthorblockA{\IEEEauthorrefmark{1}German Research Center for Artificial Intelligence (DFKI)\\Trippstadter Street 122,  Kaiserslautern, 67663 Germany\\
  }
\IEEEauthorblockA{\IEEEauthorrefmark{2}Rheinland-Pf\"alzische Technische Universit\"at  (RPTU) Kaiserslautern-Landau\\Building 11, Paul-Ehrlich Street, Kaiserslautern, 67663 Germany\\
 }
}
\maketitle

\begin{abstract}
Cell-free massive multi-input multi-output (MIMO)  has recently gained a lot of attention due to its high potential in sixth-generation (6G) wireless systems. The goal of this paper is to first present a unified modeling for massive MIMO, encompassing both cellular and cell-free architectures with a variable number of antennas per access point. We derive signal transmission models and achievable spectral efficiency in both the downlink and uplink using zero-forcing and maximal-ratio schemes. We also provide performance comparisons in terms of per-user and sum spectral efficiency. 
\end{abstract}

\section{Introduction}

Cell-free massive multi-input multi-output (MIMO) \cite{Ref_ngo2017cellfree} has recently garnered much attention in both academia and industry due to its high potential for 6G and beyond \cite{Ref_jiang2021road}. There are no cells or cell boundaries, which are essential in conventional cellular communication networks \cite{Ref_jiang2024TextBook}. Instead, a multitude of distributed access points (APs) simultaneously serve a relatively smaller user population over the same time-frequency resource. This approach ensures uniform quality of service for all users, effectively addressing the issue of under-served areas commonly encountered at the edges of conventional cellular networks \cite{Ref_jiang2023cellfree}. 

Various aspects of cell-free massive MIMO, including resource allocation \cite{Ref_buzzi2020usercentric}, power control \cite{Ref_nayebi2017precoding}, opportunistic transmission \cite{Ref_jiang2022opportunistic}, pilot assignment \cite{Ref_zeng2021pilot}, energy efficiency \cite{Ref_ngo2018total}, channel aging \cite{Ref_jiang2021impactcellfree}, backhaul constraints \cite{Ref_masoumi2020performance}, deep learning-aided methods \cite{Ref_jiang2022deep}, multi-carrier transmission \cite{Ref_jiang2021cellfree},  and scalability \cite{Ref_bjornson2020scalable}, have been studied. Nevertheless, the existing literature lacks a direct and comprehensive comparison between cellular and cell-free massive MIMO systems, with the exception of the works conducted by Yang and Marzetta in \cite{Ref_yang2018energy} and by Björnson and Sanguinetti in \cite{Ref_bjornson2019cellfree, Ref_bjornson2020making}. The former considers maximal-ratio precoding in the downlink of cell-free massive MIMO with single-antenna APs, while the latter focuses on the uplink transmission of cell-free massive MIMO.
Furthermore, prior research on cell-free massive MIMO predominantly focuses on a system model featuring either a single antenna or a fixed number of antennas per AP, leaving an unaddressed gap regarding modeling for cell-free systems that accommodate a variable number of antennas per AP.

To fill this gap, this paper aims to provide a unified modeling for massive MIMO with a flexible number of antennas per AP. This model covers both the cellular case, if all antennas are co-located at a single base station (BS), and the cell-free case, where antennas are distributed over multiple APs.  Both the downlink and uplink transmission with the aid of zero-forcing and maximal-ratio schemes are considered. Performance comparisons in terms of per-user and sum spectral efficiency  (SE) are given to provide some insightful hints on the design of efficient massive MIMO architecture. 

The main innovations of this paper are:
\begin{itemize}
    \item We provide unified modeling for massive MIMO with a variable number of antennas per AP. It is applicable for either cellular massive MIMO where all antennas are co-located at a single BS or cell-free massive MIMO with single-antenna or multi-antenna APs.
    \item We describe the signal transmission models for zero-forcing and maximal-ratio approaches in both downlink and uplink and derive their achievable SE accordingly.
    \item We numerically compare cellular and cell-free architecture in diverse settings (different precoding, detection, power control, and channel knowledge) to give insights on how to design an efficient massive MIMO system.
    \item We provide a simple power control scheme for cellular massive MIMO, achieving optimal max-min performance with low complexity. 
\end{itemize}

\begin{figure*}[!t]
    \centering
    \includegraphics[width=0.96\textwidth]{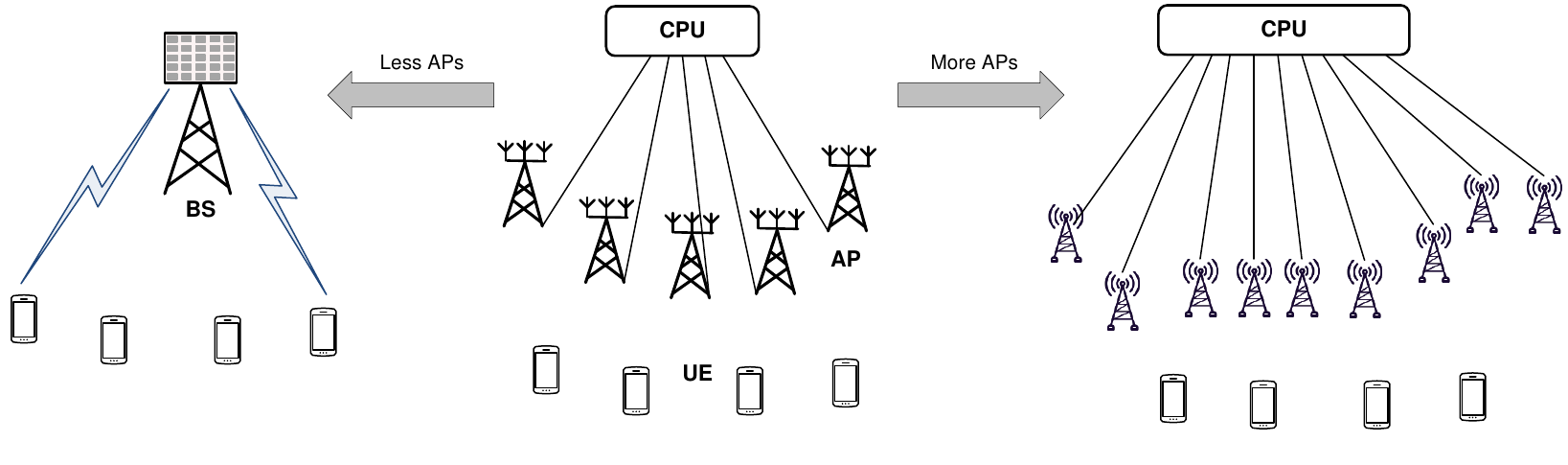}
    \caption{The unified model for cellular/cell-free massive MIMO, where $M$ antennas are distributed over $N_{AP}$ APs. If $N_{AP}=1$, it stands for cellular massive MIMO relying on a single base station, as the leftmost diagram shows, while typical cell-free massive MIMO with $N_{AP}=M$ distributed single-antenna APs is shown in the rightmost diagram.  }
    \label{fig:SystemModel}
\end{figure*}

\section{System Model}
We build a generalized model for cellular/cell-free massive MIMO, where $M$ antennas serve a few $K\ll M$ single-antenna user equipment (UEs) over an intended coverage area. These service antennas are distributed over $N_{AP}$ APs, where $1\leqslant N_{AP} \leqslant M$, and each AP is equipped with $N_{t}$ antennas. Going beyond previous works, the number of antennas per AP in our model is flexibly \textit{variable}, ranging from $ N_t=1 $ to $ N_t=M $.  This generalized model covers two special cases -- cellular massive MIMO relying on a single (i.e., $N_{AP}=1$) base station (BS), and typical cell-free massive MIMO with $N_{AP}=M$ distributed single-antenna APs, as shown in \figurename \ref{fig:SystemModel}.  A central process unit (CPU) controls all APs through a fronthaul network in the cell-free architecture, allowing it to act as a collocated antenna array. 

The channel coefficient between antenna $m$, $\forall m=1,\ldots,M$ and UE $k$, $\forall k=1,\ldots,K$ is modeled as a circularly symmetric complex Gaussian random variable, i.e., $g_{mk}\in \mathcal{CN}(0,\beta_{mk})$, where $\beta_{mk}$ stands for large-scale fading including path loss and shadowing. To minimize the significant overhead of downlink pilots, which scales with the number of service antennas, time-division duplex (TDD) is applied. Under the assumption of block fading, each coherent interval is divided into three phases: uplink training, uplink data transmission, and downlink data transmission. During uplink training, UEs transmit orthogonal pilot sequences to acquire instantaneous channel state information (CSI). Unlike multi-cell systems, pilot contamination is avoidable by increasing the length of pilot sequences. Hence, we can neglect it for simplicity.  Conducting minimum mean-square error (MMSE) estimation \cite{Ref_ngo2017cellfree}, we get $\hat{g}_{mk}\in \mathcal{CN}(0,\alpha_{mk})$ with $\alpha_{mk}=\frac{p_u\beta_{mk}^2}{p_u \beta_{mk} + \sigma_n^2}$ and the estimation error $\tilde{g}_{mk}  = g_{mk} - \hat{g}_{mk}\in \mathcal{CN}(0,\beta_{mk}-\alpha_{mk})$, where $p_u$ and $\sigma_n^2$ denote the power of UE transmitter and additive noise, respectively.
\newtheorem*{remark}{Remark}
\begin{remark}
For multi-antenna APs, large-scale fading between user $k$ and any antenna of the same AP $q$, $q=1,\ldots, N_{AP}$ is assumed to be identical. We have $\beta_{mk}=\beta_{qk}$ and $\alpha_{mk}=\alpha_{qk}$, $\forall m=(q-1)N_t{+}1,(q-1)N_t{+}2,\ldots,qN_t$. 
\end{remark}

\section{Unified Signal Models}

In the uplink data transmission, all UEs simultaneously transmit their signals towards the APs, where UE $k$ sends $\sqrt{\eta_k}x_k$. The covariance matrix of the transmit vector $\textbf{x}=[x_1,\ldots,x_K]^T$ satisfies $\mathbb{E}[\textbf{x}\textbf{x}^H]=\mathbf{I}_K$ and the power coefficient $0\leqslant \eta_k \leqslant 1$.  The network side performs linear detection to recover the transmitted symbols, resulting in 
\begin{equation} \label{eqnUPLINKmodel} \nonumber
    \textbf{y} = \textbf{A} \biggl(\sqrt{p_u} \textbf{G} \textbf{D}_\eta \textbf{x} + \textbf{n}\biggr)= \textbf{A} \left(\sqrt{p_u} \sum_{k=1}^K\textbf{g}_k \sqrt{\eta_k}x_k + \textbf{n}\right),
\end{equation}
where $\textbf{A}$ is an $K\times M$ linear detector, channel matrix $\left[\textbf{G}\right]_{mk}=g_{mk}$, $\textbf{g}_k$ is the channel signature for user $k$ (i.e., the $k^{th}$ column of $\textbf{G}$), $\textbf{D}_\eta = \mathrm{diag}([\eta_1,\ldots,\eta_K])$, and the receiver noise $\textbf{n}=[n_1,\ldots,n_M]^T\in \mathcal{CN}(\mathbf{0},\sigma^2_n\mathbf{I}_M)$.
Decomposing \eqref{eqnUPLINKmodel} yields the $k^{th}$ element:
\begin{align} \label{massiveMIMO:MFsoftestimateUL} \nonumber
    y_k &= \textbf{a}_k \left(\sqrt{p_u} \textbf{G} \textbf{D}_\eta \textbf{x} + \textbf{n}\right)\\
    &=\underbrace{ \sqrt{p_u \eta_k} \textbf{a}_k \textbf{g}_k  x_k }_{desired\:signal} + \underbrace{\sqrt{p_u}\sum_{i=1,i\neq k}^K \textbf{a}_{k} \textbf{g}_{i} \sqrt{\eta_{i}} x_{i}}_{inter-user\:interference}+\underbrace{\textbf{a}_{k}\textbf{n}}_{noise},
\end{align} where $\textbf{a}_k\in \mathbb{C}^{1\times M}$ is the $k^{th}$ row of $\textbf{A}$.

In the downlink, the network side spatially multiplexes the information-bearing symbols $\textbf{u}=[u_1,\ldots,u_K]^T$, where  $\mathbb{E}[\textbf{u}\textbf{u}^H]=\mathbf{I}_K$,  through precoding. Write $\textbf{B}$ to denote the $M\times K$ preceding matrix with entries $\left[\textbf{B}\right]_{mk}=\sqrt{\eta_{mk}}b_{mk}$, where $\eta_{mk}$ represents the power coefficient for the $k^{th}$ user at antenna $m$ while $b_{mk}$ is the precoding coefficient.  Given the per-antenna power constraint $p_d$ and noise vector $\textbf{w}=[w_1,\ldots,w_K]^T\in \mathcal{CN}(\mathbf{0},\sigma^2_n\mathbf{I}_K)$, the received symbols for all users are collectively expressed by
\begin{equation}
    \label{eqn:RxSignal}\mathbf{r} =\sqrt{p_d} \mathbf{G}^T\mathbf{B}\mathbf{u}+\mathbf{w}.
\end{equation}
Equivalently, the $k^{th}$ user has the observation of
\begin{align}\nonumber \label{EQN_downlinkModel}
    r_k &= \sqrt{p_d} \mathbf{g}_k^T\mathbf{B}\mathbf{u}+w_k\\ 
    &= \sqrt{p_d} \mathbf{g}_k^T \sum_{i=1}^K\mathbf{b}_i u_i+w_k\\\nonumber
    &=\underbrace{ \sqrt{p_d}\textbf{g}_k^T\textbf{b}_k u_k}_{desired\:signal} + \underbrace{\sqrt{p_d}\sum_{i=1,i\neq k}^K \textbf{g}_k^T \textbf{b}_{i}  u_{i}}_{inter-user\:interference}+\underbrace{w_k}_{noise},
\end{align}
where $\textbf{b}_k\in \mathbb{C}^{M\times 1}$ is the $k^{th}$ column of $\textbf{B}$.

\begin{figure*}[!t]
\begin{equation} \label{EQn_SNR_UL_MF}
    \gamma_{k} =  \frac{p_u \eta_k N_t^2\left(\sum_{q=1}^{N_{AP}}  \alpha_{qk}  \right)^2}
    {p_u N_t \sum_{i=1}^{K}  \eta_{i}  \sum_{q=1}^{N_{AP}}  \alpha_{qk}\beta_{qi}- p_u \eta_k N_t \sum_{q=1}^{N_{AP}} \alpha_{qk}^2 +\sigma^2_nN_t \sum_{q=1}^{N_{AP}}  \alpha_{qk}    }.
\end{equation}
\end{figure*}

\section{Linear Detection in Uplink}
Compared to maximum likelihood, linear detection is attractive due to its low complexity while achieving good performance. Two linear algorithms are typically applied for the uplink detection of massive MIMO \cite{Ref_jiang20226GCH9}. 
\subsection{Matched Filtering}
The philosophy behind matched filtering (MF), a.k.a. maximum-ratio combining, is to amplify the desired signal as much as possible while disregarding inter-user interference.  An MF detector is given by $\textbf{A}^{mf}=\hat{\textbf{G}}^H$ or $\textbf{a}^{mf}_k=\hat{\textbf{g}}_k^H$, where $\hat{\textbf{G}}$ is the matrix of channel estimates, i.e., $[\hat{\textbf{G}}]_{mk}=\hat{g}_{mk}$, and $\hat{\textbf{g}}_k$ is the $k^{th}$ column of $\hat{\textbf{G}}$. Substituting $\textbf{a}_k^{mf}=\hat{\textbf{g}}_k^H$ into \eqref{massiveMIMO:MFsoftestimateUL} yields 
\begin{align}  \nonumber \label{Eqn_UlCBFuserK}
    y_k &= \sqrt{p_u \eta_k} \hat{\textbf{g}}_k^H \textbf{g}_k  x_k  + \sqrt{p_u}\sum_{i=1,i\neq k}^K \hat{\textbf{g}}_k^H \textbf{g}_{i} \sqrt{\eta_{i}} x_{i}+\hat{\textbf{g}}_k^H\textbf{n}\\ \nonumber
        &= \underbrace{\sqrt{p_u \eta_k} \|\hat{\textbf{g}}_k\|^2  x_k}_{\mathcal{S}_0:\:desired\:signal}  + \underbrace{\sqrt{p_u \eta_k}  \hat{\textbf{g}}_k^H \tilde{\textbf{g}}_k  x_k}_{\mathcal{I}_1:\:channel\:estimation\:error} \\ &+\underbrace{\sqrt{p_u}\sum_{i=1,i\neq k}^K \hat{\textbf{g}}_k^H \textbf{g}_{i} \sqrt{\eta_{i}} x_{i}}_{\mathcal{I}_2}+\underbrace{\hat{\textbf{g}}_k^H\textbf{n}}_{\mathcal{I}_3},
\end{align} applying $\textbf{g}_k=\tilde{\textbf{g}}_k+\hat{\textbf{g}}_k$.

Distinct architectures of massive MIMO raise different levels of CSI availability. To be specific, the BS in cellular massive MIMO knows $\hat{\textbf{G}}$, as same as the CPU in cell-free massive MIMO when AP $m$ delivers $\hat{g}_{mk}$, $\forall k$ via the fronthaul network or the CPU conducts centralized estimation. Due to high signaling overhead, it is also possible that the CPU only knows the channel statistics. These two options correspond to:

\paragraph{Full CSI Knowledge}
Given $\hat{\textbf{g}}_k$, $\forall k$, the achievable SE of user $k$ is lower bounded by $R_k= \log(1+\gamma_k)$, where the instantaneous effective signal-to-interference-plus-noise ratio (SINR) is given by \eqref{EQn_SNR_UL_MF}.

\begin{IEEEproof}
The derivation details refer to Appendix $A$.
\end{IEEEproof}

\paragraph{Only Channel Statistics} Without the knowledge of CSI,  the CPU in a cell-free system has to detect the received signals based on $\alpha_{mk}$, $\forall m, k$. Thus, \eqref{Eqn_UlCBFuserK} is rewritten as
\begin{align}  \nonumber
    y_k &= \sqrt{p_u \eta_k} \mathbb{E}\left[\left\|\hat{\textbf{g}}_k\right\|^2\right]  x_k+\underbrace{\sqrt{p_u \eta_k} \left(\|\hat{\textbf{g}}_k\|^2 -\mathbb{E}\left[\|\hat{\textbf{g}}_k\|^2\right]\right)  x_k}_{channel\:uncertainty\:error} \\
    &+ \sqrt{p_u \eta_k}  \hat{\textbf{g}}_k^H \tilde{\textbf{g}}_k  x_k +\sqrt{p_u}\sum_{i=1,i\neq k}^K \hat{\textbf{g}}_k^H \textbf{g}_{i} \sqrt{\eta_{i}} x_{i}+\hat{\textbf{g}}_k^H\textbf{n},
\end{align}
where an additional loss due to \textit{channel uncertainty} is imposed. 
Since 
\begin{equation}
    \mathbb{E}\left(\left|\|\hat{\textbf{g}}_k\|^2 -\mathbb{E}[\|\hat{\textbf{g}}_k\|^2]\right|^2\right)=\mathrm{Var}(\|\hat{\textbf{g}}_k\|^2)= \sum_{m=1}^M \alpha_{mk}^2,
\end{equation}
we obtain the effective SINR as
\begin{equation} \label{eqn:SNR_UL_CBF_Cstats}
    \gamma_{k} =  \frac{p_u \eta_k N_t^2\left(\sum_{q=1}^{N_{AP}}  \alpha_{qk}  \right)^2}
    {p_u N_t \sum_{i=1}^{K}  \eta_{i}  \sum_{q=1}^{N_{AP}}  \alpha_{qk}\beta_{qi} +\sigma^2_nN_t \sum_{q=1}^{N_{AP}}  \alpha_{qk}    }.
\end{equation}

\begin{remark}
By setting $N_{AP}=1$,  \eqref{EQn_SNR_UL_MF} and \eqref{eqn:SNR_UL_CBF_Cstats} are applied for massive MIMO.  In addition, we get the performance of a general cell-free massive MIMO with $1< N_{AP} \leqslant M$ single- or multi-antenna APs.
\end{remark}

\subsection{Zero-Forcing Detection}
Instead of maximizing the strength of the desired signal, the principle of ZF is to minimize inter-user interference. The ZF detector is the pseudo inverse of the channel matrix, i.e., $\textbf{A}_{zf}=(\hat{\textbf{G}}^H\hat{\textbf{G}})^{-1}\hat{\textbf{G}}^H$. This method needs full CSI, which is easy for the cellular setup but it imposes high overhead signaling in a cell-free system. Substituting $\textbf{a}_{k}^{zf}$, the $k^{th}$ row of $\textbf{A}_{zf}$, into \eqref{massiveMIMO:MFsoftestimateUL} yields  
\begin{align}  \label{EQn_UL_ZFD_perUserrate} \nonumber
    y_k &= \sqrt{p_u \eta_k}  x_k  + \sqrt{p_u \eta_k} \textbf{a}_k^{zf} \tilde{\textbf{g}}_k  x_k\\ \nonumber &+\sqrt{p_u}\sum_{i=1, i\neq k}^K \sqrt{\eta_i} \textbf{a}_k^{zf} \tilde{\textbf{g}}_i  x_i+\textbf{a}_k^{zf}\textbf{n}\\
    &= \sqrt{p_u \eta_k}  x_k  +\underbrace{\sqrt{p_u}\sum_{i=1}^K \sqrt{\eta_i} \textbf{a}_k^{zf} \tilde{\textbf{g}}_i  x_i}_{channel\:estimation\:error}+\textbf{a}_k^{zf}\textbf{n}.
\end{align}
The SINR of ZF detection can be given by
\begin{equation} \label{EqN_UL_ZFD_perUserRateForm}
    \gamma_{k}=  \frac{p_u \eta_k}
    {\phi_k\left(p_u \sum_{i=1}^K\eta_i N_t  \sum_{q=1}^{N_{AP}} (\beta_{qi}-\alpha_{qi} )  + \sigma^2_n \right)}.
\end{equation}
\begin{IEEEproof}
The derivation details refer to Appendix $B$.
\end{IEEEproof}

\begin{figure*}[!tbph]
\centerline{
\subfloat[]{
\includegraphics[width=0.44\textwidth]{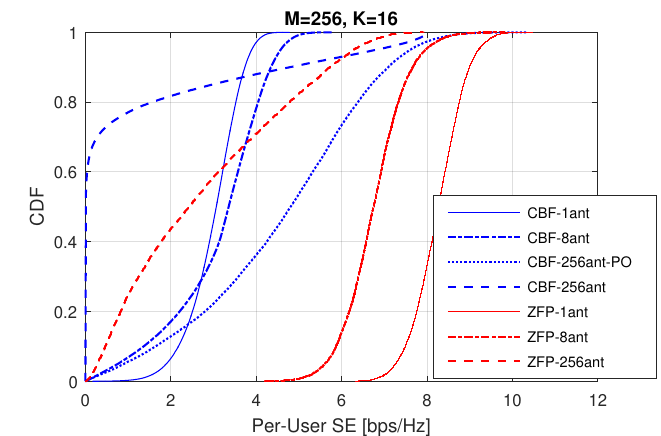}
\label{fig:result1} 
}
\hspace{20mm}
\subfloat[]{
\includegraphics[width=0.44\textwidth]{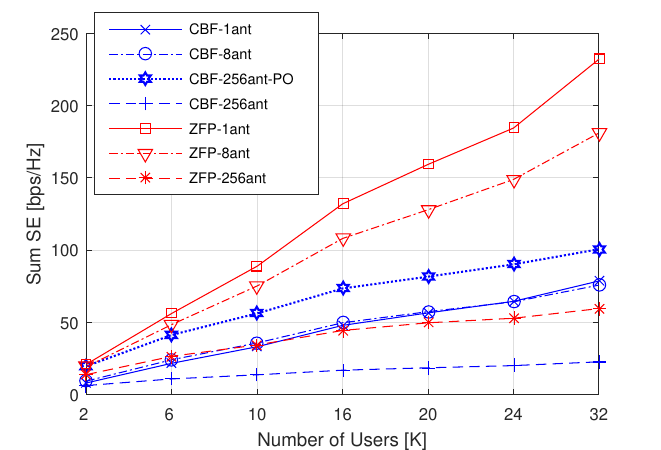}
\label{fig:result2}
}
}
\hspace{15mm}
 \caption{Performance comparison in the downlink of massive MIMO systems: (a) the CDF curves in terms of per-user SE using ZFP and CBF, where $M=256$ antennas serve $K=16$ users; and (b) the achievable sum SE in terms of the number of users when $M=256$.    }
 \label{fig:result}
\end{figure*}

\section{Linear Precoding in Downlink}
In the downlink, two precoding methods, i.e., conjugate beamforming (CBF) and zero-forcing precoding (ZFP), are applied to spatially multiplex the information-bearing symbols intended for $K$ terminals.

\subsubsection{Conjugate Beamforming}

As the MF in the uplink, conjugate beamforming aims to maximize the desired signal. The precoding matrix is given by $\left[\textbf{B}\right]_{mk}=\sqrt{\eta_{mk}}\hat{g}^*_{mk}$. 
Thus, \eqref{EQN_downlinkModel} is decomposed into an element-wise form as
\begin{align}\nonumber 
    r_k  &= \sqrt{p_d} \sum_{m=1}^M \sqrt{\eta_{mk}} \hat{g}_{mk}\hat{g}_{mk}^* u_k + \sqrt{p_d} \sum_{m=1}^M \sqrt{\eta_{mk}} \tilde{g}_{mk}\hat{g}_{mk}^* u_k\\
         &+\sqrt{p_d}\sum_{i=1,i\neq k}^K  \sum_{m=1}^M \sqrt{\eta_{mi}} g_{mk}\hat{g}_{mi}^*  u_i  + w_k.
\end{align}
Each user only knows channel statistics $ \mathbb{E} \left[ \left |  \hat{g}_{mk} \right | ^2\right]=\alpha_{mk}$ rather than channel estimate $\hat{g}_{mk}$ since there are no downlink pilots in \textit{either cellular or cell-free massive MIMO}. The power control is generally decided by $\beta_{qk}$, we have $\eta_{mk}=\eta_{qk}$ for $m=(q-1)N_t{+}1,\ldots,qN_t$. Using similar manipulations as the derivation of \eqref{EQn_SNR_UL_MF}, we obtain the effective SINR as
\begin{equation}  \label{EQN_DLCBFPerUserRate}
    \gamma_{k}=  \frac{p_d N_t^2\left(\sum_{q=1}^{N_{AP}} \sqrt{\eta_{qk}} \alpha_{qk}  \right)^2}
    {\sigma^2_n+p_d N_t \sum_{q=1}^{N_{AP}} \beta_{qk} \sum_{i=1}^{K}  \eta_{qi} \alpha_{qi} }.
\end{equation}

\begin{remark}
Let $N_{AP}=M$,  \eqref{EQN_DLCBFPerUserRate} is aligned with (27) of \cite{Ref_ngo2017cellfree}, which indicates the performance of cell-free massive MIMO with single-antenna APs. Hence, the correctness of \eqref{EQN_DLCBFPerUserRate} justifies.
\end{remark}

\paragraph{Zero-Forcing Precoding}
This precoding at the network side eliminates the inter-user interference at the UE receiver. The precoding matrix is the pseudo inverse of the channel matrix, i.e., $\textbf{B}=\hat{\textbf{G}}^*(\hat{\textbf{G}}^T\hat{\textbf{G}}^*)^{-1}\odot \mathbf{E}$, where $\odot$ means the Hadamard product and $\left[\textbf{E}\right]_{mk}=\sqrt{\eta_{mk}}$. As proved by \cite{Ref_nayebi2017precoding}, it is necessary to have $\eta_{1k}=\eta_{2k}=\cdots=\eta_{Mk}$, $\forall k$ to keep $\hat{\textbf{G}}^T\textbf{B}$ orthogonal such that inter-user interference is eliminated. We have $\eta_{mk}=\eta_{k}$, $\forall m$ and therefore $\textbf{B}=\hat{\textbf{G}}^*(\hat{\textbf{G}}^T\hat{\textbf{G}}^*)^{-1} \mathbf{D}$, where $\textbf{D} = \mathrm{diag}([\eta_1,\ldots,\eta_K])$. 
In \eqref{EQN_downlinkModel},  the inter-user interference $\sum_{i=1,i\neq k}^K \textbf{g}_k^T \textbf{b}_{i}  u_{i}=0$,   we have
\begin{align}\nonumber 
    r_k &=  \sqrt{p_d}\textbf{g}_k^T\textbf{b}_k u_k +w_k\\ \nonumber
        &=  \sqrt{p_d}\hat{\textbf{g}}_k^T\textbf{b}_k u_k +\sqrt{p_d}\tilde{\textbf{g}}_k^T\textbf{b}_k u_k+w_k\\
        &=  \sqrt{p_d \eta_k} u_k +\sqrt{p_d}\tilde{\textbf{g}}_k^T\textbf{b}_k u_k+w_k.
\end{align}
The effective SINR of user $k$ is expressed by
\begin{equation}  
    \gamma_k = \frac{p_d\eta_k}{\sigma_n^2 + p_d\sum_{i=1}^K \eta_i \chi_i^k}.
\end{equation}
where $\chi_{i}^k$ denotes the $i^{th}$ diagonal element of the $K\times K$ matrix dedicated to user $k$:
\begin{equation} \label{EQNZFP_Constatnt}
    \mathbb{E}\biggl[ \left(\hat{\mathbf{G}}\hat{\mathbf{G}}^H\right)^{-1} \hat{\mathbf{G}}\mathbb{E}\left[\tilde{\textbf{g}}_k^H\tilde{\textbf{g}}_k\right]  \hat{\mathbf{G}}^H\left(\hat{\mathbf{G}}\hat{\mathbf{G}}^H\right)^{-1}\biggr],
\end{equation}
and $\mathbb{E}\left[\tilde{\textbf{g}}_k^H\tilde{\textbf{g}}_k\right]$ is a diagonal matrix with the $k^{th}$ diagonal element equaling to $\beta_{mk}-\alpha_{mk}$. 

Although \eqref{EQNZFP_Constatnt} is general, as done in  \cite{Ref_nayebi2017precoding}, we can also get a dedicated SE formula for the cellular case. As $\mathbb{E}[\tilde{\textbf{g}}_k^H\tilde{\textbf{g}}_k]$ is simplified to $(\beta_{k}-\alpha_{k})\mathbf{I}_M$, \eqref{EQNZFP_Constatnt} equals to $(\beta_{k}-\alpha_{k})\mathbb{E}[ (\hat{\mathbf{G}}\hat{\mathbf{G}}^H)^{-1} ]$.
Thus,  the effective SINR of user $k$ in cellular massive MIMO is obtained as
\begin{equation}  
    \gamma_k = \frac{p_d\eta_k}{\sigma_n^2 + p_d (\beta_{k}-\alpha_{k}) \sum_{i=1}^K \eta_i \phi_i}.
\end{equation}

\section{Numerical Results}
The performance of cell-free and cellular massive MIMO is comparatively evaluated in terms of per-user and sum spectral efficiency. 
Let a total of $M=256$ antennas serve users over a square area of $1\times 1\mathrm{km^2}$. Large-scale fading is given by $\beta_{mk}=10^\frac{\mathcal{L}_{mk}+\mathcal{X}_{mk}}{10}$ with shadowing $\mathcal{X}_{mk}\sim \mathcal{N}(0,\sigma_{sd}^2)$, where $\sigma_{sd}=8\mathrm{dB}$, and path loss, as calculated by the COST-Hata model  \cite{Ref_ngo2017cellfree}:
\begin{equation} 
    \mathcal{L}_{mk}= \begin{cases}
-L_0-35\log_{10}(d_{mk}), &  d_{mk}>d_1 \\
-L_0-10\log_{10}(d_1^{1.5}d_{mk}^2), &  d_0<d_{mk}\leq d_1 \\
-L_0-10\log_{10}(d_1^{1.5}d_0^2), &  d_{mk}\leq d_0
\end{cases},
\end{equation}
where the three-slope breakpoints  take values $d_0=10\mathrm{m}$ and $d_1=50\mathrm{m}$ while $L_0=140.72\mathrm{dB}$ in terms of 
\begin{IEEEeqnarray}{ll}
 L_0=46.3&+33.9\log_{10}\left(f_c\right)-13.82\log_{10}\left(h_{AP}\right)\\ \nonumber
 &-\left[1.1\log_{10}(f_c)-0.7\right]h_{UE}+1.56\log_{10}\left(f_c\right)-0.8
\end{IEEEeqnarray}
with carrier frequency $f_c=1.9\mathrm{GHz}$, the height of AP antenna $h_{AP}=15\mathrm{m}$, and the height of UE $h_{UE}=1.65\mathrm{m}$. Per-antenna and UE power constraints are set to $p_d=200\mathrm{mW}$ and $p_u=100\mathrm{mW}$, respectively.  The white noise power density equals $-174\mathrm{dBm/Hz}$ with a noise figure of $9\mathrm{dB}$, and the signal bandwidth is set to $5\mathrm{MHz}$.

The uplink transmission is carried out in a distributed manner, it is reasonable that each UE simply use a full-power strategy without global power control. In the downlink, unfortunately, the optimal max-min power-control schemes in cell-free massive MIMO, both with CBF and ZFP, are too computationally complex for practical use. As suggested by \cite{Ref_nayebi2017precoding}, we adopt sub-optimal schemes with low complexity.  To be specific, in ZFP, $\eta_{1}=\ldots=\eta_{K}=\left( \max_m  \sum_{k=1}^{K} \delta_{km} \right)^{-1}$, where $\boldsymbol \delta_m= [\delta_{1m},\ldots,\delta_{Km}]^T=\mathrm{diag}(\mathbb{E}[  (\hat{\mathbf{G}}\hat{\mathbf{G}}^H)^{-1}    \hat{\mathbf{g}}_m \hat{\mathbf{g}}_m^H   \hat{\mathbf{G}}\hat{\mathbf{G}}^H)^{-1} ])$ and $\hat{\mathbf{g}}_m$ denotes the $m^{th}$ column of $\hat{\mathbf{G}}$. In CBF, the APs utilize the full-power strategy, i.e.,  $\eta_{m}=(\sum_{k=1}^{K} \alpha_{mk} )^{-1}$, $\forall m$. Thanks to the identical channel statistics over all antennas in cellular massive MIMO, we can provide a simple but optimal power-control scheme for CBF,  as derived in Appendix $C$.

\begin{figure}[!t]
    \centering
    \includegraphics[width=0.46\textwidth]{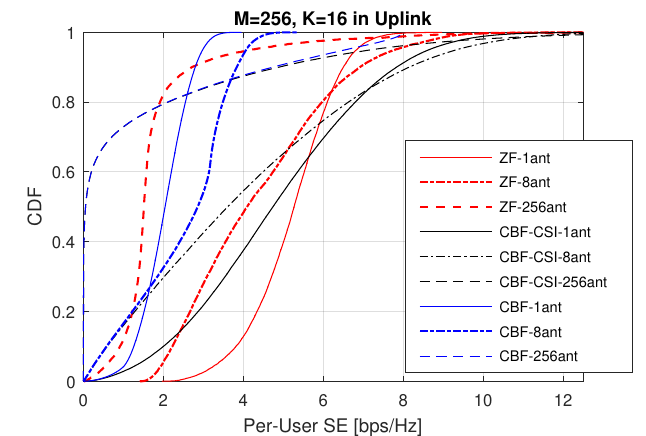}
    \caption{Comparison of the CDF in terms of per-user SE in the uplink of massive MIMO systems using ZF and MF detection, where $M=256$ antennas serve $K=16$ users.  }
    \label{fig:results3}
\end{figure}

\figurename \ref{fig:result}a provides a comparison with respect to cumulative distribution functions (CDFs) of per-user SE in the downlink. It includes three representative antenna configurations:  cellular with a 256-antenna BS, and cell-free with single-antenna or 8-antenna APs. In general, the performance of ZFP is better than that of CBF under the same antenna configuration, while a higher level of antenna distribution brings larger per-user SE. To be specific, the $5\%$-likely per-user SE for CBF with one and eight antennas are around $1.89\mathrm{bps/Hz}$ and $0.76\mathrm{bps/Hz}$, respectively, in comparison with $7.26\mathrm{bps/Hz}$ and $5.61\mathrm{bps/Hz}$ of ZFP. In cellular massive MIMO, although the performance of ZFP is far better than that of CBF, the power-control scheme given in Appendix $C$ can bring a remarkable gain, outperforming ZFP. It achieves $5\%$-likely per-user SE of $0.91\mathrm{bps/Hz}$, which is larger than $0.28\mathrm{bps/Hz}$ obtained by ZFP in cellular massive MIMO, justifying its advantage. 

In \figurename \ref{fig:result}b, we illustrate sum rates of different schemes with respect to a variety of user numbers when $K=2, 6, 10, 16, 20, 24$, and $32$. Ignoring the overhead of orthogonal uplink pilots, which is proportional to the number of users, more users can generate higher system capacity due to the multi-user gain. Also, ZFP is better than CBF while cell-free configuration outperforms cellular ones. Using the given power-control scheme, however, CBF can still perform much better than ZFP in cellular massive MIMO.
Moreover, \figurename \ref{fig:results3} compares per-user SE of different schemes in the uplink.  The particularity of the uplink is that the receiver is able to conduct maximal filtering based on full CSI, rather than only channel statistics. With the aid of channel estimates, MF achieves comparable performance as zero-forcing detection. If there is no CSI available, the performance of MF substantially degrades. It reveals the significance of the CSI availability in a massive MIMO system.

\section{Conclusion}
This paper provides unified modeling of massive MIMO systems with a variable number of antennas per AP in both downlink and uplink. It applies to both cellular and cell-free architectures with maximal-ratio and zero-forcing methods. Through extensive performance comparison and analyses, we highlight 1) the significance of channel knowledge, by which zero forcing gains advantages; 2) the importance of power control, where cellular configuration is not necessarily worse; and 3) the utility of near-far effect due to antenna distribution. It is our hope that the insights presented in this paper will contribute to the design of an efficient massive MIMO system for 6G and beyond.

\appendices
\section{Derivation of Per-User SE with MF}
The terms $\mathcal{S}_0$, $\mathcal{I}_1$, $\mathcal{I}_2$, and $\mathcal{I}_3$ in \eqref{Eqn_UlCBFuserK} are mutually uncorrelated. According to \cite{Ref_hassibi2003howmuch}, the worst-case noise for mutual information is Gaussian additive noise with the variance equalling to the variance of $\mathcal{I}_1+\mathcal{I}_2+\mathcal{I}_3$. 
Thus, the uplink achievable rate for user $k$ is lower bounded by $R_k= \log(1+\gamma_k)$,
where
\begin{align}  \label{cfmmimo:formularSNR}
    \gamma_k  = \frac{\mathbb{E}\left[|\mathcal{S}_0|^2\right]}{\mathbb{E}\left[|\mathcal{I}_1+\mathcal{I}_2+\mathcal{I}_3|^2\right]}
         = \frac{\mathbb{E}\left[|\mathcal{S}_0|^2\right]}{\mathbb{E}\left[|\mathcal{I}_1|^2\right]+\mathbb{E}\left[|\mathcal{I}_2|^2\right]+\mathbb{E}\left[|\mathcal{I}_3|^2\right]}
\end{align}
with 
\begin{align}  \label{APPEQ1}
    \mathbb{E}\left[|\mathcal{S}_0|^2\right] & = p_u\eta_k N_t^2\left( \sum_{q=1}^{N_{AP}} \alpha_{qk} \right)^2\\ \label{APPEQ2}
    \mathbb{E}\left[|\mathcal{I}_1|^2\right] & = p_u\eta_kN_t\sum_{q=1}^{N_{AP}} (\beta_{qk}-\alpha_{qk})\alpha_{qk}\\ \label{APPEQ3}
    \mathbb{E}\left[|\mathcal{I}_2|^2\right] & = p_u N_t \sum_{i=1,i\neq k}^K \eta_i \sum_{q=1}^{N_{AP}}  \beta_{qi} \alpha_{qk}\\  \label{APPEQ4}
    \mathbb{E}\left[|\mathcal{I}_3|^2\right] & = \sigma_n^2N_t\sum_{q=1}^{N_{AP}} \alpha_{qk}
\end{align}
Substituting the above terms into \eqref{cfmmimo:formularSNR}, yields \eqref{EQn_SNR_UL_MF}.

\section{Derivation of Per-User SE with ZF Detection}
It is not difficult to get
\begin{align}
    \mathbb{E}\left[\left\|\textbf{a}_{k}^{zf}\right\|^2\right]=\mathbb{E}\left[\left[\textbf{A}_{zf}\textbf{A}_{zf}^H \right]_{kk} \right]=\mathbb{E}\left[\left[(\hat{\textbf{G}}^H\hat{\textbf{G}})^{-1} \right]_{kk} \right],
\end{align}
where $[\cdot]_{kk}$ denotes the $k^{th}$ diagonal element of a matrix. Let $\phi_k=\mathbb{E}[[(\hat{\textbf{G}}^H\hat{\textbf{G}})^{-1} ]_{kk} ]$ for simple notation, see \eqref{EQn_UL_ZFD_perUserrate}, we further derive that
\begin{align}  
    \mathbb{E}\left[\left|\sum_{i=1}^K \sqrt{\eta_i} \textbf{a}_k^{zf} \tilde{\textbf{g}}_i  x_i\right|^2\right] & = \phi_k \sum_{i=1}^K \eta_i N_t \sum_{q=1}^{N_{AP}} (\beta_{qi}-\alpha_{qi})   \\ 
    \mathbb{E}\left[ \left| \textbf{a}_k^{zf}\textbf{n} \right|^2 \right] & = \phi_k \sigma_n^2
\end{align}
Thus, we get \eqref{EqN_UL_ZFD_perUserRateForm}.

\section{Power Optimization for Cellular Massive MIMO}
For massive MIMO, \eqref{EQN_DLCBFPerUserRate} is transferred to 
\begin{equation}  \label{EQNAPP01}
    \gamma_{k}=  \frac{p_d M^2\eta_{k} \alpha_{k}^2}
    {\sigma^2_n+p_d M \beta_{k} \sum_{i=1}^{K}  \eta_{i} \alpha_{i} }.
\end{equation}
The power constraint is $\sum_{k=1}^{K}  \eta_{k} \alpha_{k} \leqslant 1$. Assume $\sum_{k=1}^{K}  \eta_{k} \alpha_{k} \neq 1$, we can get a common factor $\theta>1$, satisfying still $\sum_{k=1}^{K}  \theta \eta_{k} \alpha_{k} \leqslant 1$. Thus, \eqref{EQNAPP01} becomes
\begin{align} \nonumber
    \gamma_{k} &=  \frac{p_d M^2  \theta \eta_{k} \alpha_{k}^2}
    {\sigma^2_n+p_d M \beta_{k} \sum_{i=1}^{K} \theta  \eta_{i} \alpha_{i} }=\frac{p_d M^2   \eta_{k} \alpha_{k}^2}
    {\sigma^2_n/\theta+p_d M \beta_{k} \sum_{i=1}^{K}   \eta_{i} \alpha_{i} },
\end{align}  
implying that collectively increasing power can benefit all users simultaneously, when $\sum_{k=1}^{K}  \eta_{k} \alpha_{k} < 1$. Therefore, $\sum_{k=1}^{K}  \eta_{k} \alpha_{k} = 1$ is optimal. 
Thus, \eqref{EQNAPP01} becomes 
\begin{equation}  
    \gamma_{k}=  \frac{p_d M^2\eta_{k} \alpha_{k}^2}
    {\sigma^2_n+p_d M \beta_{k} }.
\end{equation}
The max-min power optimization can be expressed by
\begin{equation}  
\begin{aligned} \label{eqnIRS:optimizationMRTvector}
\max_{\eta_k,\:\forall k} \min_{k}\quad &  \gamma_{k} \\
\textrm{s.t.} \quad & \sum_{k=1}^{K}  \eta_{k} \alpha_{k} = 1,
\end{aligned}
\end{equation}
which is equivalent to 
\begin{equation}  
\begin{aligned}
\max_{\eta_k,\:\forall k} \quad &  t \\
\textrm{s.t.} \quad & \gamma_{k} \geqslant t, \forall k\\
\quad & \sum_{k=1}^{K}  \eta_{k} \alpha_{k} = 1.
\end{aligned}
\end{equation}
Setting $\gamma_k=t$, $\forall k$ results in 
\begin{equation}  \label{EQNappterm}
 \eta_{k} \alpha_{k}   =  t \left( \frac
    {\sigma^2_n+p_d M \beta_{k} } {p_d M^2 \alpha_{k}}\right)
\end{equation}
Since $\sum_{k=1}^{K}  \eta_{k} \alpha_{k} = 1$, we have 
\begin{equation}  
 \sum_{k=1}^{K}\eta_{k} \alpha_{k}   =  t \sum_{k=1}^{K}\left( \frac
    {\sigma^2_n+p_d M \beta_{k} } {p_d M^2 \alpha_{k}}\right)=1.
\end{equation}
Thus, 
\begin{equation}  
 t   =  \frac {1}{\sum_{k=1}^{K}\left( \frac
    {\sigma^2_n+p_d M \beta_{k} } {p_d M^2 \alpha_{k}}\right)}.
\end{equation}
Substituting $t$ into \eqref{EQNappterm}, we get the optimal power coefficients 
\begin{equation}  
 \eta_{k}    =  \frac {\left( \frac
    {\sigma^2_n+p_d M \beta_{k} } {p_d M^2 \alpha_{k}}\right)}{\alpha_{k} \sum_{i=1}^{K}\left( \frac
    {\sigma^2_n+p_d M \beta_{i} } {p_d M^2 \alpha_{i}}\right)}.
\end{equation}

\bibliographystyle{IEEEtran}
\bibliography{IEEEabrv,Ref_COML}

\begin{thebibliography}{10}
\providecommand{\url}[1]{#1}
\csname url@samestyle\endcsname
\providecommand{\newblock}{\relax}
\providecommand{\bibinfo}[2]{#2}
\providecommand{\BIBentrySTDinterwordspacing}{\spaceskip=0pt\relax}
\providecommand{\BIBentryALTinterwordstretchfactor}{4}
\providecommand{\BIBentryALTinterwordspacing}{\spaceskip=\fontdimen2\font plus
\BIBentryALTinterwordstretchfactor\fontdimen3\font minus \fontdimen4\font\relax}
\providecommand{\BIBforeignlanguage}[2]{{%
\expandafter\ifx\csname l@#1\endcsname\relax
\typeout{** WARNING: IEEEtran.bst: No hyphenation pattern has been}%
\typeout{** loaded for the language `#1'. Using the pattern for}%
\typeout{** the default language instead.}%
\else
\language=\csname l@#1\endcsname
\fi
#2}}
\providecommand{\BIBdecl}{\relax}
\BIBdecl

\bibitem{Ref_ngo2017cellfree}
H.~Q. Ngo \emph{et~al.}, ``Cell-free massive {MIMO} versus small cells,'' \emph{{IEEE} Trans. Wireless Commun.}, vol.~16, no.~3, pp. 1834--1850, Mar. 2017.

\bibitem{Ref_jiang2021road}
W.~Jiang \emph{et~al.}, ``The road towards {6G}: A comprehensive survey,'' \emph{IEEE Open J. Commun. Society}, vol.~2, pp. 334--366, Feb. 2021.

\bibitem{Ref_jiang2024TextBook}
W.~Jiang and B.~Han, \emph{Cellular Communication Networks and Standards: The Evolution from {1G} to {6G}}.\hskip 1em plus 0.5em minus 0.4em\relax Cham, Switzerland: Springer, 2024.

\bibitem{Ref_jiang2023cellfree}
W.~Jiang and H.~D. Schotten, ``Cell-edge performance booster in {6G}: Cell-free massive {MIMO} vs. reconfigurable intelligent surface,'' in \emph{Proc. {IEEE} Eur. Conf. on Netw. and Commun. (EUCNC)}, Gothenburg, Sweden, Jun. 2023, pp. 1--6.

\bibitem{Ref_buzzi2020usercentric}
S.~Buzzi \emph{et~al.}, ``User-centric {5G} cellular networks: Resource allocation and comparison with the cell-free massive {MIMO} approach,'' \emph{{IEEE} Trans. Wireless Commun.}, vol.~19, no.~2, pp. 1250--1264, Feb. 2020.

\bibitem{Ref_nayebi2017precoding}
E.~Nayebi \emph{et~al.}, ``Precoding and power optimization in cell-free massive {MIMO} systems,'' \emph{{IEEE} Trans. Wireless Commun.}, vol.~16, no.~7, pp. 4445--4459, Jul. 2017.

\bibitem{Ref_jiang2022opportunistic}
W.~Jiang and H.~Schotten, ``Opportunistic {AP} selection in cell-free massive {MIMO-OFDM} systems,'' in \emph{Proc. 2022 IEEE 95th Veh. Techno. Conf. (VTC2022-Spring)}, Helsinki, Finland, Jun. 2022.

\bibitem{Ref_zeng2021pilot}
W.~Zeng \emph{et~al.}, ``Pilot assignment for cell-free massive {MIMO} systems using a weighted graphic framework,'' \emph{{IEEE} Trans. Veh. Technol.}, pp. 6190 -- 6194, Jun. 2021.

\bibitem{Ref_ngo2018total}
H.~Q. Ngo \emph{et~al.}, ``On the total energy efficiency of cell-free massive {MIMO},'' \emph{IEEE Trans. Green Commun. Netw.}, vol.~2, no.~1, pp. 25--39, Mar. 2018.

\bibitem{Ref_jiang2021impactcellfree}
W.~Jiang and H.~Schotten, ``Impact of channel aging on zero-forcing precoding in cell-free massive {MIMO} systems,'' \emph{{IEEE} Commun. Lett.}, vol.~25, no.~9, pp. 3114 -- 3118, Sep. 2021.

\bibitem{Ref_masoumi2020performance}
H.~Masoumi and M.~J. Emadi, ``Performance analysis of cell-free massive {MIMO} system with limited fronthaul capacity and hardware impairments,'' \emph{{IEEE} Trans. Wireless Commun.}, vol.~19, no.~2, pp. 1038--1052, Feb. 2020.

\bibitem{Ref_jiang2022deep}
W.~Jiang and H.~Schotten, ``Deep learning-aided delay-tolerant zero-forcing precoding in cell-free massive {MIMO},'' in \emph{Proc. 2022 IEEE 96th Veh. Techno. Conf. (VTC2022-Fall)}, London, UK, Sep. 2022.

\bibitem{Ref_jiang2021cellfree}
W.~Jiang and H.~D. Schotten, ``Cell-free massive {MIMO-OFDM} transmission over frequency-selective fading channels,'' \emph{{IEEE} Commun. Lett.}, vol.~25, no.~8, pp. 2718 -- 2722, Aug. 2021.

\bibitem{Ref_bjornson2020scalable}
E.~Björnson and L.~Sanguinetti, ``Scalable cell-free massive {MIMO} systems,'' \emph{{IEEE} Trans. Commun.}, vol.~68, no.~7, pp. 4247--4261, Jul. 2020.

\bibitem{Ref_yang2018energy}
H.~Yang and T.~L. Marzetta, ``Energy efficiency of massive {MIMO}: Cell-free vs. cellular,'' in \emph{Proc. {IEEE} 87th Veh. Techno. Conf. (VTC-Spring)}, Porto, Portugal, Jun. 2018.

\bibitem{Ref_bjornson2019cellfree}
E.~Björnson and L.~Sanguinetti, ``Cell-free versus cellular massive {MIMO}: What processing is needed for cell-free to win?'' in \emph{Proc. {IEEE} 20th Int. Workshop Signal Process. Advances in Wireless Commun. (SPAWC)}, Cannes, France, Jul. 2019.

\bibitem{Ref_bjornson2020making}
------, ``Making cell-free massive {MIMO} competitive with {MMSE} processing and centralized implementation,'' \emph{{IEEE} Trans. Wireless Commun.}, vol.~19, no.~1, pp. 77 -- 90, Jan. 2020.

\bibitem{Ref_jiang20226GCH9}
W.~Jiang and F.-L. Luo, \emph{6G Key Technologies: A Comprehensive Guide}.\hskip 1em plus 0.5em minus 0.4em\relax New York, USA: IEEE Press \& Wiley, 2023, ch.~9.

\bibitem{Ref_hassibi2003howmuch}
B.~Hassibi and B.~Hochwald, ``How much training is needed in multiple-antenna wireless links?'' \emph{{IEEE} Trans. Inf. Theory}, vol.~49, no.~4, pp. 951 -- 963, Apr. 2003.

\end{thebibliography}

\end{document}